\documentclass{article}

\usepackage{PRIMEarxiv}

\usepackage[utf8]{inputenc} 
\usepackage[T1]{fontenc}    
\usepackage{hyperref}       
\usepackage{url}            
\usepackage{booktabs}       
\usepackage{amsfonts}       
\usepackage{nicefrac}       
\usepackage{microtype}      
\usepackage{lipsum}
\usepackage{fancyhdr}       
\usepackage{graphicx}       
\usepackage{amsmath}
\usepackage{bm}
\usepackage[ruled,vlined,linesnumbered]{algorithm2e}
\graphicspath{{media/}}     
\usepackage{comment}

\title{PS-WL: A Probability-Sensitive Wear Leveling scheme for SSD array scaling
}

\author{
    Shuhang Xu, Yunfei Gu, Linhui Liu, Chentao Wu\\
    Shanghai Jiao Tong University\\
    Shanghai, China
}

\begin{document}
\maketitle

\begin{abstract}

As flash-based Solid State Drive (SSD) arrays become essential to modern data centers, scaling these arrays to meet explosive data growth is a frequent and critical operation. However, the conventional wear-leveling (WL) paradigm applied during scaling suffers from a fundamental flaw: it ignores the non-linear relationship between wear and failure probability, potentially pushing the most vulnerable, aged disks towards premature failure. To address this critical issue at its root, we propose the Probability-Sensitive Wear Leveling (PS-WL) scheme, which shifts the optimization goal from balancing wear to directly balancing failure risk. At its core, PS-WL introduces an "effective lifetime" model derived from a realistic failure probability to more accurately assess disk lifetime. This model guides a PID controller for wear leveling operation, with a conservative zone minimizes performance overhead by restricting warm data migration. Comprehensive simulations validate the superiority of PS-WL over state-of-the-art methods. The results demonstrate that our approach significantly reduces performance overhead while, most critically, consistently and effectively lowering the aggregated array failure risk across diverse system configurations and workloads. This proves that by directly optimizing for reliability, PS-WL builds a scalable storage system that is, by design, fundamentally safer, more efficient, and more stable.

\end{abstract}


\section{Introduction}

With the rapid development of flash memory manufacturing technology and continuous cost optimization, solid-state drives (SSDs) have gradually replaced traditional hard disk drives (HDDs) as the preferred storage medium for modern data centers and enterprise-level storage systems due to their outstanding read and write performance, high reliability, low power consumption, and excellent shock resistance\cite{4694025}\cite{10.1145/1982185.1982266}. Under the global digital transformation wave, the volume of data is increasing at an unprecedented rate. According to statistics, the total global data volume is expected to reach 394 ZB by 2028 \cite{taylor2024amount}. From the PB-level text corpora required for training large language models (LLMs) to the TB-level sensor data generated by autonomous vehicles every hour \cite{Gehrig2024}, and to the vast storage infrastructure built by major cloud service providers such as Google Cloud Storage \cite{10.5555/2930583.2930589}, Microsoft Azure \cite{azure_premium_storage}, and Amazon AWS\cite{8030603} to support the business operations of millions of users, all these have continuously posed challenges to the capacity and performance of storage systems. To address these challenges and ensure high data availability, the use of Redundant Array of Independent Disks (RAID) and SSD array scaling have become routine operations in data center maintenance.

However, many traditional array scaling schemes were originally designed for hard disk drives (HDDs), and when applied to SSD arrays, they often cause a series of problems due to neglecting the unique physical characteristics of SSDs, such as limited Program/Erase (P/E) cycles, Write Amplification (WA), Wear Leveling (WL), and Garbage Collection (GC).
In the context of array scaling, the newly extended disks, with their extremely low initial wear levels, are immediately identified by the wear-leveling algorithm as the ideal write target disk. This results in a highly centralized I/O workloads on the new disks. Consequently, it not only causes I/O congestion on the new disks, significantly degrading the array's overall response performance, but also leads to an excessive accumulation of hot data on the new disks. This introduces additional I/O overhead and unnecessary write amplification.. 

To solve these challenges, researchers have proposed advanced schemes like SWANS \cite{10.1145/2756555}, Lazy-WL \cite{9556067}, and Ada-WL \cite{10527393}. These methods attempt to smooth the lifetime gap by monitoring workloads and managing data migration. However, a significant limitation persists: most existing schemes use P/E cycles as the sole measure of disk lifetime, failing to account for the accelerated failure probability inherent to SSDs. The correlation between an SSD's P/E cycles and its failure probability is highly non-linear; after a certain threshold, the failure probability rises sharply \cite{7105166}. This non-uniform aging can cause disks to fail at different times, which complicates RAID management and can trigger slow, performance-degrading data rebuilds, or even cause direct data loss.

Therefore, a more advanced wear-leveling scheme is needed—one that not only balances the I/O load but also proactively manages the reliability risk of the entire array. To this end, this paper proposes PS-WL (Probability-Sensitive Wear Leveling). The core idea of PS-WL is to replace the traditional P/E count metric with an "effective lifetime" model that incorporates the non-linear failure probability of SSDs. By doing so, PS-WL aims to achieve array-level wear leveling while minimizing the aggregated failure probability of the entire array, leading to smarter, more durable load balancing that protects data and ensures long-term system stability.

The main contributions of this paper are as follows:

We propose a new wear-leveling scheme, PS-WL. Its key innovation is an effective lifetime model based on SSD failure probability. This model provides a more accurate assessment of disk health, shifting the optimization goal from simply balancing P/E cycles to reducing the entire array's aggregated failure probability. This helps avoid overusing disks that are at a high risk of failure during scaling.
To test and evaluate the PS-WL scheme, we designed and implemented a high-fidelity SSD array simulator. The simulator can model the SSD's P/E process, wear characteristics, and failure probability changes in detail, providing a solid experimental basis for evaluating the performance and reliability of our scheme and related research.
Through extensive simulations, we show that PS-WL outperforms existing advanced schemes like Lazy-WL and Ada-WL. In array scaling scenarios, especially with large wear differences between old and new disks, PS-WL shows significant advantages in reducing overall array failure probability, balancing I/O loads, and improving long-term system stability.
The rest of this paper is organized as follows: We introduce the related work in Section II. In Section III, we explain the design and implementation details of the PS-WL scheme. In Section IV, we present the detailed experimental setup and result analysis. Finally, We draw our conclusion in Section V.

\section{Background and Related Work}

This chapter aims to establish the theoretical and technical foundations for the proposed PS-WL scheme. We will first introduce the background. Following this, we provide a detailed analysis of the Solid State Drive failure probability model. Based on this, we examine representative RAID scaling schemes and inter-disk wear-leveling schemes. Finally, we present the motivation for this work.

\subsection{NAND technology and SSD}

A Solid State Drive (SSD) is a storage device based on NAND flash memory. Unlike traditional Hard Disk Drives (HDDs), which rely on mechanical rotation and magnetic head seeking, SSDs read and write data directly using electrical signals. This allows them to achieve microsecond-level access latency, extremely high IOPS (Input/Output Operations Per Second) and lower power consumption. However, the physical properties of NAND flash also introduce unique challenges: its storage cells have a limited number of Program/Erase (P/E) cycles. To manage these limitations, SSDs use the Flash Translation Layer (FTL). The FTL is responsible for three critical tasks: address mapping (translating the host's logical addresses to the physical addresses of the flash memory), Garbage Collection (GC, which reclaims invalid data space created by "out-of-place updates"), and Wear Leveling (WL, which ensures that all flash blocks are worn as evenly as possible). Together, these mechanisms determine the performance and lifespan of an SSD.

The storage core of a SSD is the NAND Flash memory. Its concept is based on a metal oxide semiconductor device with a Floating Gate (FG) electrically isolated by means of a tunnel oxide.
Raw bit errors in NAND Flash memory primarily stem from the technology's inherent physical limitations, namely its finite endurance and imperfect data retention. An error is registered whenever a memory cell's threshold voltage shifts away from its intended, programmed level.
During the early stages of an SSD's usage, although P/E cycles increase, the damage to the flash cells is relatively minor. The Raw Bit Error Rate (RBER) grows slowly, and the built-in Error Correction Code (ECC) can easily correct these errors. During this phase, the disk's failure probability increases at a slow rate. However, once the P/E cycle is accumulated beyond a certain threshold, multiple physical degradation mechanisms (such as tunnel oxide degradation, random telegraph noise and cell-to-cell interference)  causing the RBER to grow exponentially. As soon as the RBER growth rate surpasses the correction capability of the ECC, uncorrectable errors occur more frequently, marking the SSD's accelerated failure probability. At this point, even a small increase in P/E cycles can cause a sharp rise in failure probability, rapidly pushing the disk toward its reliability tipping point.

This "failure probability acceleration phenomenon"\cite{7105166}\cite{298621}\cite{10.5555/1960475.1960493} highlights the inadequacy of balancing P/E cycles alone. The reliability gap between two highly worn disks can be far greater than their P/E count difference suggests. This critical insight forms the basis for our proposed "effective lifetime" metric, which will be detailed in Section III.

\subsection{Existing RAID Scaling Approaches}

The following are several representative scaling methods:

\begin{itemize}
    \item \textit{Round-Robin (RR)}: RR is a basic and simple-to-implement scaling approach \cite{10.1145/1162628.1162631}
    \cite{10.1145/1227835.1227838}. 
    It migrates existing data blocks in a round-robin sequential manner across the entire array, including the new disks, and is widely used in various RAID configurations. While this method achieves optimal data layout balance, it comes at the cost of an extremely high data migration rate, involving nearly all data and resulting in massive I/O overhead and parity update calculations.
    \item \textit{FastScale (FS)}: FS is a lightweight disk array scaling method designed for RAID 0 \cite{10.5555/1960475.1960486}. It selectively migrates data from the old disks to the new disks, avoiding data migration between the old disks. This significantly reduces the amount of data migration. However, the limitation is that it cannot be directly applied to RAID levels that require parity.
    \item \textit{Global Stripe-based Redistribution (GSR)}: GSR is a scaling scheme optimized for RAID 5 \cite{6337607}. It operates from the perspective of the entire array's stripe layout, using intelligent partitioning and reorganization to minimize data migration and the overhead of parity updates.
    \item \textit{Stripe-based Data Migration (SDM)}: SDM is a scaling method designed for RAID 6 \cite{6337790}. It optimizes data migration according to the target parity layout, effectively reducing data migration and complex dual-parity calculations while ensuring uniform data distribution.
\end{itemize}

\subsection{Existing Wear Leveling Approaches}
Inter-disk wear leveling techniques aim to balance the average WL among different SSDs in array scaling scenario. 

\begin{itemize}
    \item \textit{SWANS}: SWANS (Smoothing Wear Across N SSDs)\cite{10.1145/2756555} focus on write balancing in array. It monitors the write workload on each disk and adjust write distribution across SSD array, migrate hot data blocks to less active disks, and optimizes the wear chase by increasing the write operations.

    \item \textit{EDM}: Endurance-aware Data Migration\cite{6877310} prolong SSD endurance and boost system performance by minimizing the write amplification inherent to the migration process. The method incorporates a hot-data-first policy, giving priority to the migration of hot data with the highest write frequency to quickly balance the load. Furthermore, a supplementary cold-data-first policy is used to migrate infrequently accessed cold data.
       
    \item \textit{Lazy-WL}: Lazy-WL (Lazy Wear-Leveling)\cite{9556067} recognizes the problem of excessive writing to new disks in the initial phase of scaling. It proposes a "lazy" strategy, centered on establishing a "Conservative Zone" that restricts the migration of hot data to new disks. It also introduces a PID controller to monitor the lifetime standard deviation among the original and extended SSDs, ensure a smooth and controlled wear catch-up process.
    \item \textit{Ada-WL}: Ada-WL (Adaptive Wear-Leveling)\cite{10527393} improves upon Lazy-WL by using a machine learning model to dynamically adjust the conservative zone range and employing adaptive control theory to optimize the PID parameters, making it adaptable to more complex and dynamic workloads.
\end{itemize}

\subsection{Summary and Motivation}

In summary, while existing scaling and wear-leveling schemes address performance bottlenecks, they predominantly rely on P/E counts as the primary wear metric. As detailed in Section II, this approach overlooks the non-linear, accelerated failure probability of aging SSDs. This oversight can lead to strategies that, while balancing P/E counts, inadvertently increase the overall array failure risk. Our work is motivated by the need for a reliability-aware wear-leveling scheme that directly addresses this gap.
In this way, PS-WL aims to achieve a better wear leveling in terms of failure rate, which is especially critical for ensuring long-term data reliability and system stability when managing scaling scenarios that include long-serving disks.

\section{System Design of PS-WL scheme}

To address the challenges of SSD array scaling, we propose the Probability-Sensitive Wear Leveling (PS-WL) scheme. The design of PS-WL is centered on shifting the optimization goal from balancing P/E counts to balancing reliability risk. Its architecture is composed of four key modules that work in synergy:

\begin{enumerate}
    \item \textbf{Effective Lifetime Assessment}: At the core of PS-WL is a model that translates raw P/E cycles into an "effective lifetime" metric by incorporating the non-linear failure probability of SSDs. This provides a more accurate measure of disk lifetime.
    \item \textbf{Parameter Collection}: This module gathers essential static parameters (e.g., array size) and dynamic state information (e.g., P/E counts) required for the model and control logic.
    \item \textbf{Data Management and Restriction}: To minimize overhead, this module identifies data hotness to guide migration decisions and employs a conservative zone to protect warm data from inefficient migrating.
    \item \textbf{Wear Leveling Control}: A PID-based controller monitor the entire wear chase process. It intelligently determines the start timing, manages the wear catch-up speed based on the effective lifetime difference, and decides when to exit the wear chase process.
\end{enumerate}

The following sections will detail the design and implementation of each of these modules.

\subsection{Parameter Collection and Perception}

To achieve precise and adaptive wear leveling, the PS-WL scheme must first comprehensively collect key parameters of the system. These parameters includes:
the number of original SSDs in the array $k_o$,
the number of newly added SSDs for scaling $k_s$,
the number of blocks per disk $N_b$,
the number of pages per block $N_p$,
the maximum P/E cycles for a data block $L_m$, and
$t_i$: The P/E cycle count of the $i$-th SSD in the array at the current time. This is the most direct indicator of physical wear and the basis for calculating the effective lifetime.

The accurate acquisition of these parameters is a prerequisite for all subsequent calculations and decisions. 

\subsection{Effective Lifetime Model}

As established previously, traditional wear-leveling algorithms that rely solely on P/E cycle counts fail to capture the non-linear failure risk of SSDs. To overcome this, PS-WL introduces a model based on failure probability.

To accurately capture the non-linear changes in SSD failure probability, this study adopts the Log-Logistic distribution for modeling, drawing on relevant empirical research \cite{7105166}. Its S-shaped curve characteristic effectively describes the smooth transition from a low probability in the early stages to a high probability in the later stages. Given a raw P/E cycle count $t$, the cumulative failure probability of an SSD, $P_{failure}(t)$, is defined as follows:
\begin{equation}
P_{failure}(t) = \frac{1}{1 + \exp\left(-\frac{\log(t) - \mu}{\sigma}\right)}\label{eq:Pfail}
\end{equation}
where $t$ is the current raw P/E cycle count, $\mu$ is the scale parameter and $\sigma$ is the shape parameter of the log-logistic distribution respectively. These two parameters must be fitted from the lifetime testing data of a specific SSD model.

To integrate non-linear failure probability information into the wear-leveling framework, this study proposes the concept of effective lifetime. The effective lifetime $L(t)$ not only considers the raw P/E cycles but also incorporates a risk penalty term determined by the current failure probability, defined as follows:
\begin{equation}
L(t) = k \cdot t + k_p \cdot P_{failure}(t)
\end{equation}
where $t$ is the raw P/E cycle count, $P_{failure}(t)$ is the cumulative failure probability calculated by Equation \ref{eq:Pfail}, and $k$ and $k_p$ configurable penalty weight coefficient for the failure probability. A larger value of $k_p$ makes the system more "sensitive" to disks with high failure probability.
This model achieves a non-linear correction of a disk's aging status under high P/E cycles by introducing the failure probability term. All subsequent decisions in the PS-WL scheme, such as start timing, wear catch-up, and exit conditions, will be based on this effective lifetime $L(t)$, which more accurately reflects the overall health status, rather than on the simple P/E cycle count.

\subsection{Hot Data Identification and Restriction} \label{hot}

To enable a more granular quantification of data hotness within the sliding window, we introduce a multi-dimensional evaluation model based on Data Access Dynamics. This model comprehensively assesses the significance of data by analyzing it from three aspects: frequency, recency, and compactness, thereby constructing a dynamic hotness profile.

Specifically, we move beyond traditional monolithic metrics and instead examine the following three metrics:

\begin{itemize}
    \item \textbf{Access Frequency} : This indicator measures the total number of accesses to a data page within the observed sliding window.It is a direct measure of \textbf{frequency locality}, identifying pages that are consistently popular over a period.
    \item \textbf{Access Recency}: This indicator measures the elapsed time, in terms of access events, since the most recent reference to a data page. A smaller value indicates a more recent access. It is used to evaluate \textbf{temporal locality}, reflecting how recent the access is.
    \item \textbf{Access Compactness}: This indicator measures the number of intervening data accesses between the two most recent references to a specific page. It directly quantifies the reuse proximity of accesses. A smaller value signifies a tightly clustered access pattern, revealing a strong short-term correlation in the access stream.
\end{itemize}

This multi-dimensional assessment method allows for a more accurate differentiation between various types of hot data, providing a reliable basis for subsequent differentiated management. These hotness metrics are min-max normalized.

Based on the multi-dimensional hotness, the system dynamically classifies data pages into three categories and applies different migration principles:

\begin{itemize}
    \item Extremely Hot Data: After wear leveling begins, these are preferentially guided to migrate to new disks with low effective lifetimes to quickly reduce the lifetime gap using their high update frequency.
    \item Cold Data: These are completely excluded from active migration for the purpose of wear leveling to avoid invalid overhead.
    \item Warm Data: This is the largest portion of active data. To prevent them from being migrated on a large scale and inefficiently during wear leveling, PS-WL designs an conservative zone for them.
\end{itemize}

The core idea of the conservative zone is to establish a data protection area that includes warm data with relatively stable access patterns but not "extremely hot" enough to warrant immediate migration. When the system prepares to execute a data migration, if the target data page is in the conservative zone and the destination is an extended disk, the system will adopt a protection strategy based on the current degree of effective lifetime difference. If the difference is not large, the migration is canceled to prevent excessive accumulation of warm data on the extended disks.
The capacity of the conservative zone, $C$, is not fixed but is adaptively adjusted based on the system state, calculated as follows:
\begin{equation} \label{eq:Lban}
C = N_{base} \times k_{ban}
\end{equation}
where $N_{base}$ is the number of data pages in the entire array, and $k_{ban}$ is a dynamic adjustment factor that incorporates factors such as wear status and lifetime differences.

\subsection{Wear Leveling Control}
Once the PS-WL scheme is started, the core task is to perform a "wear catch-up" between the original and newly extended disks. To finely control the catch-up speed, PS-WL uses a multi-variable PID controller that combines the real-time access hotness of the disks and the difference in effective lifetime.
Based on the indicators introduced in Section \ref{hot},  the hotness of disks can be represented as a three-dimensional Dynamic Hotness Vector $\mathbf{R}(t)$:

\begin{equation} \label{eq:hotness_vector}
  \mathbf{R}(t) = [H_{\text{freq}}(t), H_{\text{rec}}(t), H_{\text{comp}}(t)]
\end{equation}

where:
\begin{itemize}
  \item $H_{\text{freq}}(t)$ is the \textbf{Frequency Hotness}, directly calculated from the Access Frequency. A higher access count results in a higher hotness value.
  \item $H_{\text{rec}}(t)$ is the \textbf{Recency Hotness},  inversely derived from the Access Recency. A more recent access (smaller recency value) corresponds to a higher hotness value.
  \item $H_{\text{comp}}(t)$ is the \textbf{Compactness Hotness},  inversely mapped from the Access Compactness. A tighter cluster of accesses (smaller compactness value) yields a higher hotness value.
\end{itemize}

Next, the PID controller calculates a theoretical hotness baseline, $u(t)$, based on the average effective lifetime difference, $e(t)$, between the original disk group and the extended disk group.

\begin{equation} \label{eq:pid_error}
e(t) = L_{o}(t) - L_{s}(t)
\end{equation}
where $L_{o}(t)$ and $L_{s}(t)$ are the average effective lifetimes of the original and extended disk groups, respectively. Based on this error, the PID controller calculates the theoretical hotness baseline $u(t)$:
\begin{equation}\label{eq:pid_output}
u(t) = K_{p}e(t) + K_{i}\sum_{i=0}^{t}e(i) + K_{d}|e(t)-e(t-T_{0})|
\end{equation}
where $K_p, K_i, K_d$ are the PID coefficients and $T_0$ is the sampling period. When the actual hotness $\mathbf{R}(t)$ is below the theoretical baseline $u(t)$, the system is inclined to approve data migrations.

To adapt to different workloads, PS-WL also implements a PID parameter self-tuning mechanism, using the ratio of wear-leveling I/O to total I/O, $J$, as the loss function for the optimization of the PID coefficients $K =[K_{p},K_{i},K_{d}]$:


\begin{equation} \label{eq:pid_tuning_sign}
K_{n+1} = K_n - \alpha_n \cdot \text{sign}(J(K_{n-1}) - J(K_n)) \cdot e_j
\end{equation}

where $\alpha$ is the learning rate, $e_j$ is the unit vector. This allows PS-WL to continuously optimize its own control performance.

To avoid unnecessary continuous adjustments, the wear-leveling operation will be paused or exited when the average effective lifetime difference between the original and extended disks decreases to a sufficiently small relative threshold.
\begin{equation} \label{eq:exit_condition}
\frac{|L_{o}(t) - L_{s}(t)|}{L_{o}(t)} < \lambda
\end{equation}
where $\lambda$ is a preset exit threshold.

\section{Experimental Evaluation and Analysis}

This chapter aims to comprehensively evaluate the effectiveness of our proposed PS-WL scheme through a series of detailed simulation experiments. We will first introduce our experimental methodology, including the experimental platform, baseline schemes, datasets, and evaluation metrics. Subsequently, we will present the performance of PS-WL under various RAID configurations and scaling scenarios, comparing it against current mainstream inter-disk wear-leveling schemes. Finally, we will conduct an in-depth analysis of the experimental results to explain the underlying reasons for the performance advantages of the PS-WL scheme.

\subsection{Methodology}
To ensure the accuracy and reproducibility of our results, we implemented an SSD emulator capable of accurately simulating the low-level behavior of NAND flash, including P/E operations, Garbage Collection (GC), and wear leveling. To align with the scheme designed in Chapter III, we further implemented support for the SSD failure probability model in the emulator, enabling it to track and calculate the effective lifetime of each simulated disk.

We select several classic scaling methods deployed by the SSD array, including Round-Robin (RR), FastScale (FS), Global Stripe-based Redistribution (GSR), and Stripe-based Data Migration (SDM). These methods are classical designs for scaling disk arrays in RAID-0, RAID-5, and RAID-6, allowing for a comprehensive evaluation of PS-WL's adaptability across different array architectures.

To evaluate the performance of our proposed \textbf{PS-WL} scheme, we conduct a comprehensive comparison against three baseline approaches: \textbf{SWANS}\cite{10.1145/2756555}, \textbf{Lazy-WL}\cite{9556067} and \textbf{EDM}\cite{6877310}, with different scaling methods in various RAID configurations. We used the Alibaba Cloud Block Storage Traces\cite{9251252}, a write-intensive, large-scale I/O trace for our experiments. 


\subsection{Experimental Results}

We conducted tests on the SSD emulator we designed for $(k_o,k_s)$ scaling scenario where $k_o$ is the number of original disk and $k_s$ is the number of scaling disk. To quantify the direct benefit of our probability-sensing model, we designed a specific experiment comparing PS-WL against other WL approach. .

\subsubsection{Lifetime Standard Deviation}

The primary objective of this section is to evaluate the effectiveness of the PS-WL scheme in achieving wear balance among drives within an SSD array. We use the standard lifetime deviation as the core metric for assessing array-level wear-leveling effectiveness. A lower lifetime standard deviation indicates that the wear-leveling strategy has effectively distributed the write workload across all disks. To facilitate a cross-scheme comparison, we represent this metric using the standard deviation of the Program/Erase (P/E) cycle counts of all disks in the array.

The experimental results demonstrate that PS-WL consistently maintains the lifetime standard deviation at an extremely low level across all scaling scenarios and workloads, significantly outperforming all comparative schemes. This directly proves the precision and effectiveness of the PS-WL strategy in tracking and rectifying wear imbalances. In contrast, SWANS, Lazy-WL, and EDM produced higher lifetime standard deviations in most cases, indicating that their wear-leveling approaches have significant deficiencies in adaptability across various scenarios.

\begin{figure}[!htb]
    \centering
    \includegraphics[width=0.7\textwidth]{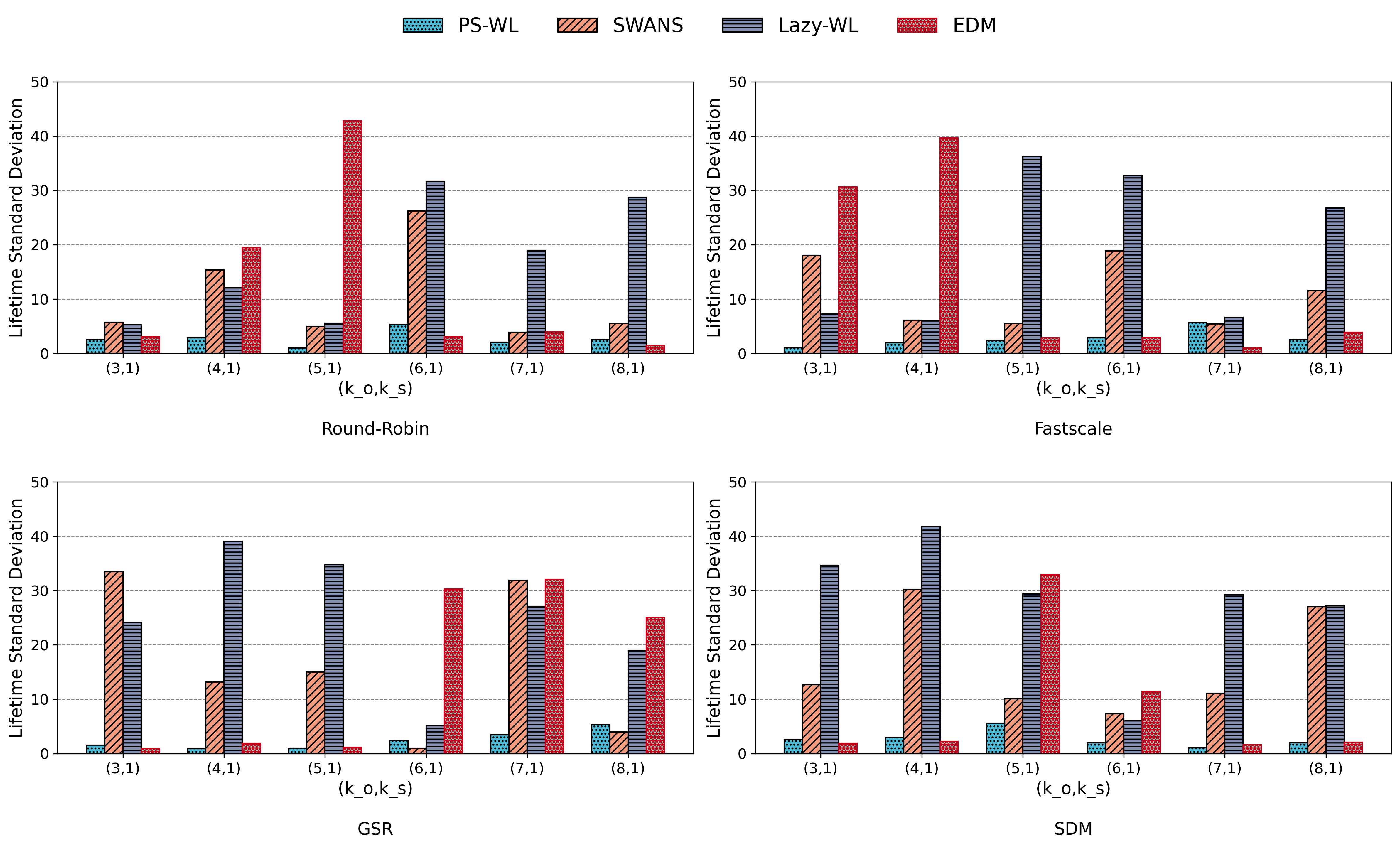}
    \caption{Lifetime Standard Deviation result when adding original disks}
    \label{fig:exp1_mu_alibaba_single}
\end{figure}

\begin{figure}[!htb]
    \centering
    \includegraphics[width=0.7\textwidth]{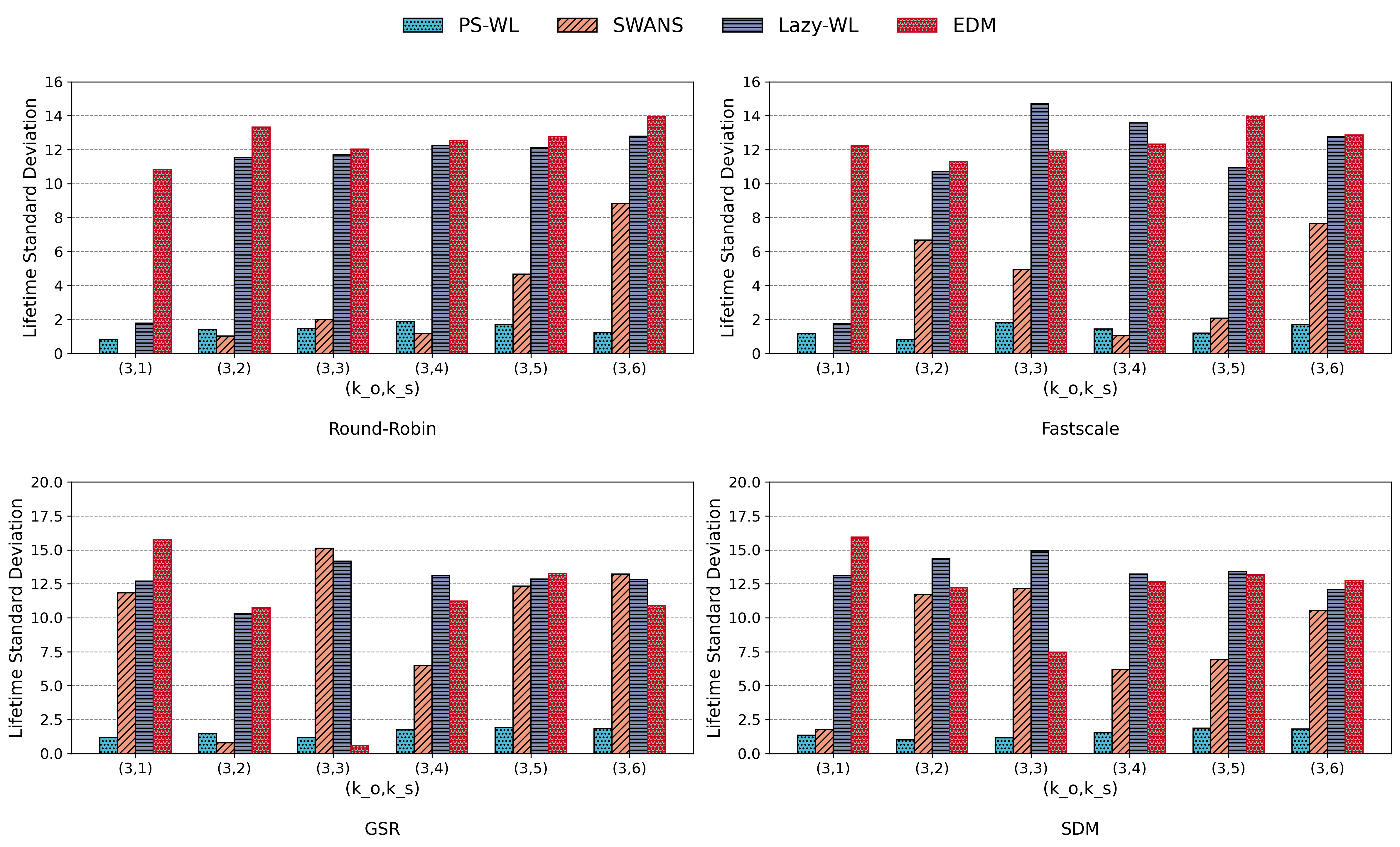}
   \caption{Lifetime Standard Deviation result when adding scaling disks}
    \label{fig:exp1_mu_alibaba_multiple}
\end{figure}

Figure \ref{fig:exp1_mu_alibaba_single} and \ref{fig:exp1_mu_alibaba_multiple} present the results for write-intensive workloads. PS-WL reduced the lifetime standard deviation by an average of approximately 79.9\%, 88.5\%, and 83.7\% compared to SWANS, Lazy-WL, and EDM, respectively.

\subsubsection{Average Response Time}

This section aims to evaluate the impact of the PS-WL scheme on system I/O performance during SSD array scaling, specifically measured by the key metric of Average Response Time. In enterprise-level storage systems, maintaining a stable and low-latency response time is crucial, particularly when executing background tasks such as data migration and wear-leveling operations. A lower average response time signifies that the scaling process causes less disruption to system performance. 

The experimental results clearly demonstrate that the PS-WL scheme exhibits superior performance across all tested combinations. Its average response time is consistently maintained at an extremely low level, and it shows exceptional stability. In contrast, the three comparative schemes—SWANS, Lazy-WL, and EDM—all induced significant performance jitter and increased latency, with their average response times being far higher than that of PS-WL.

\begin{figure}[!htb]
    \centering
    \includegraphics[width=0.7\textwidth]{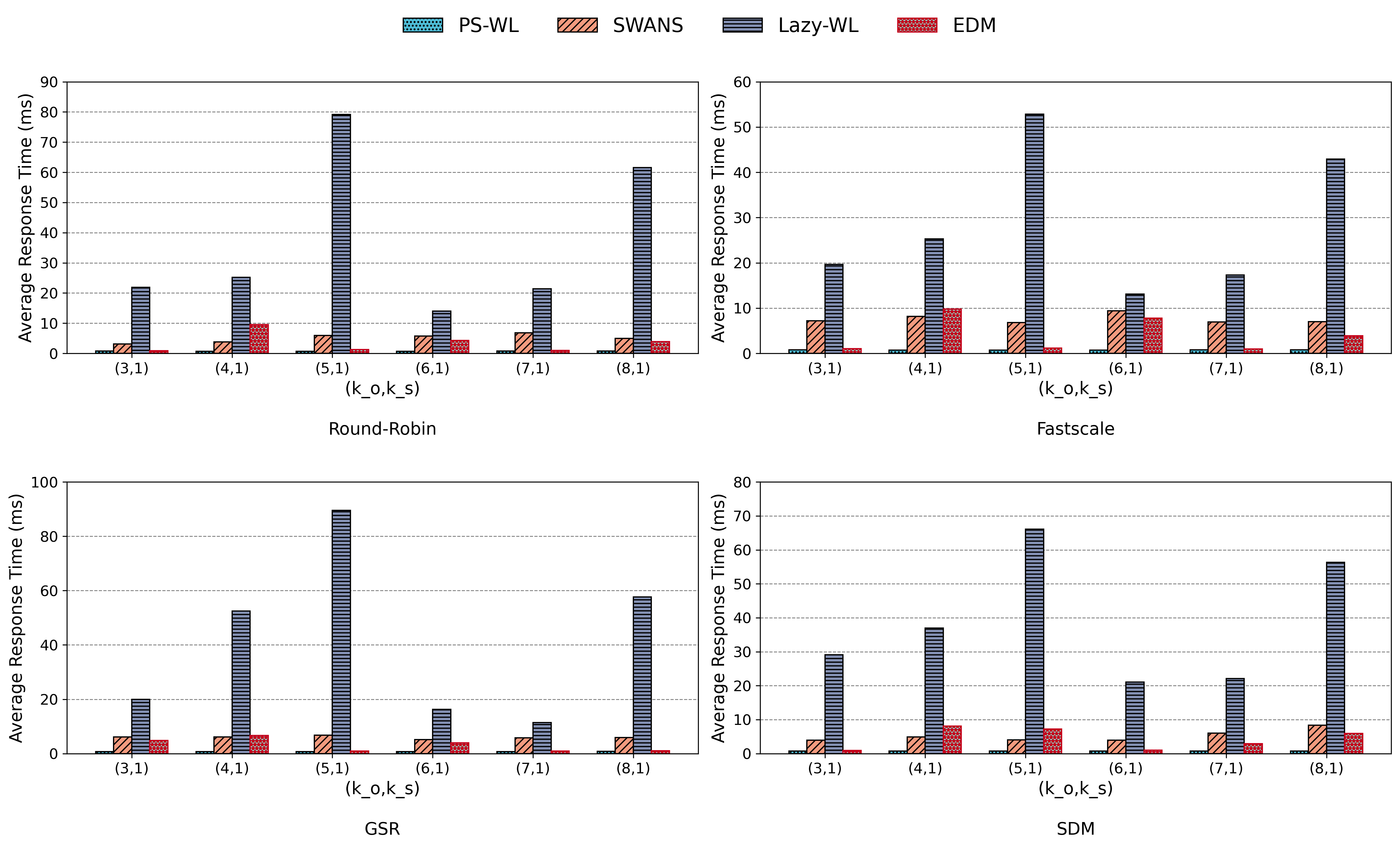}
\caption{Average response time result when adding original disks}
    \label{fig:exp1_art_alibaba_single}
\end{figure}

\begin{figure}[!htb]
    \centering
    \includegraphics[width=0.7\textwidth]{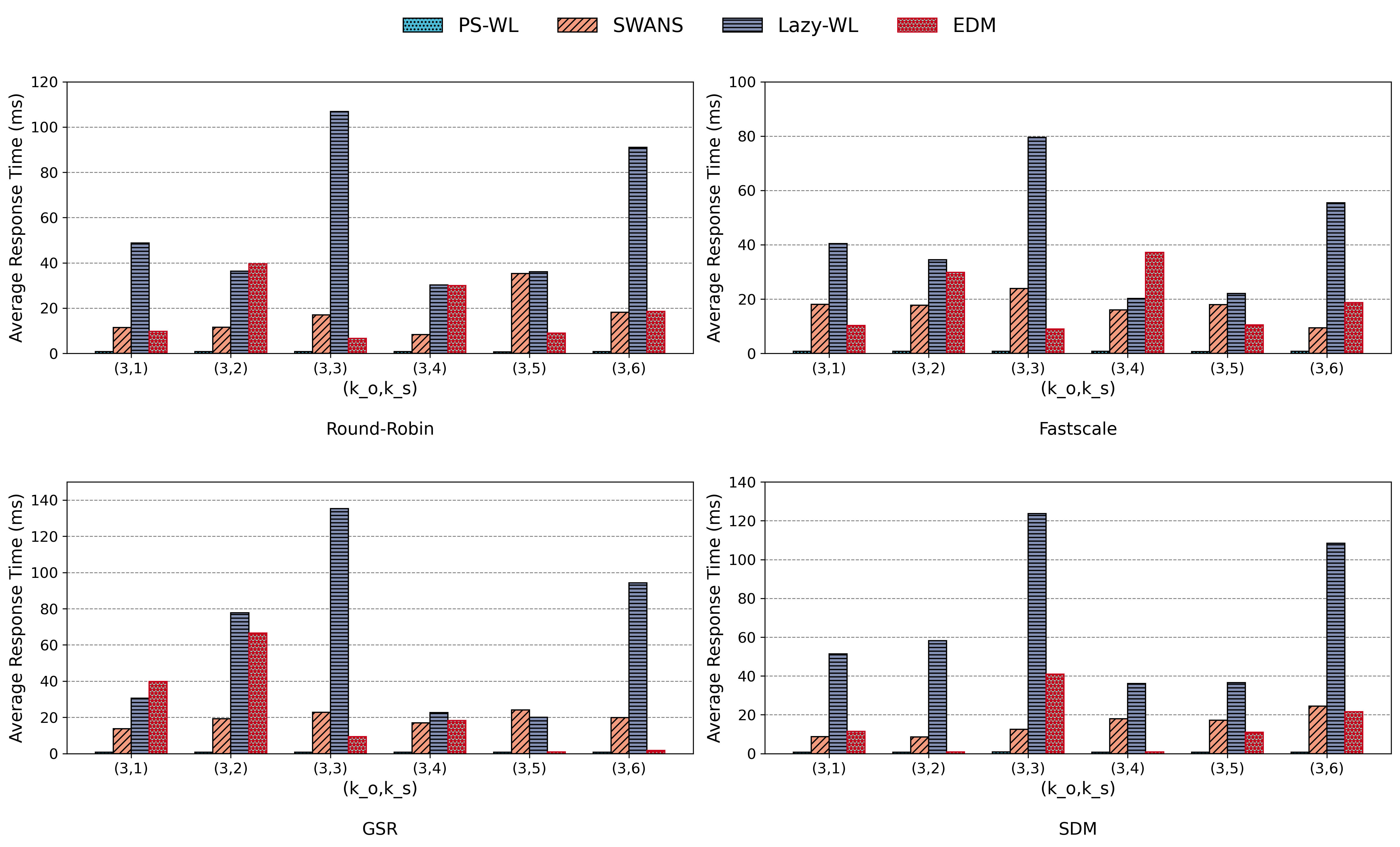}
\caption{Average response time result when adding scaling disks}
    \label{fig:exp1_art_alibaba_multiple}
\end{figure}

As shown in Figure \ref{fig:exp1_art_alibaba_single} and \ref{fig:exp1_art_alibaba_multiple}, under write-intensive workload, the effects of write amplification and garbage collection on SSDs become more pronounced. This, in turn, further amplifies the potential impact of wear-leveling operations on performance. Compared to SWANS, Lazy-WL, and EDM, the PS-WL scheme reduced the average response time by approximately 92.8\%, 97.5\%, and 86.4\%, respectively.

\subsubsection{Wear-Leveling Trigger Count}

This section analyzes the Wear-Leveling Trigger Count, a metric that records the total number of times the wear-leveling algorithm was activated to perform data migration throughout the simulation. It directly quantifies the operational overhead of the wear-leveling scheme itself. An efficient algorithm should achieve the desired balancing effect with an appropriate number of triggers and a suitable amount of data migration. Triggering too frequently can introduce unnecessary performance overhead and additional write amplification, whereas triggering too infrequently may indicate that the wear-leveling strategy is sluggish. 

\begin{figure}[!htb]
    \centering
    \includegraphics[width=0.7\textwidth]{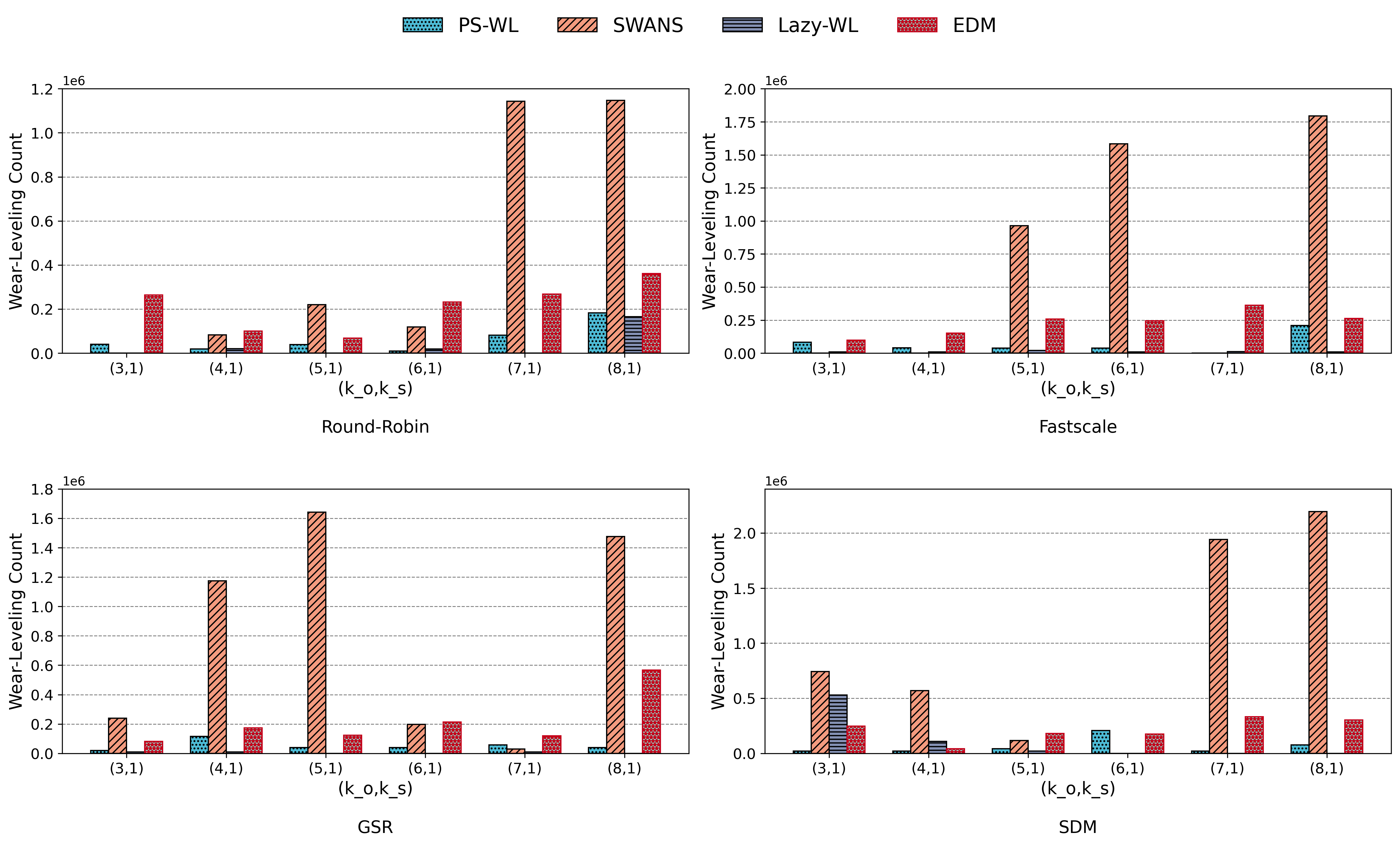}
\caption{Wear-Leveling Trigger Count result when adding original disks}
    \label{fig:exp1_wl_alibaba_single}
\end{figure}

\begin{figure}[!htb]
    \centering
    \includegraphics[width=0.7\textwidth]{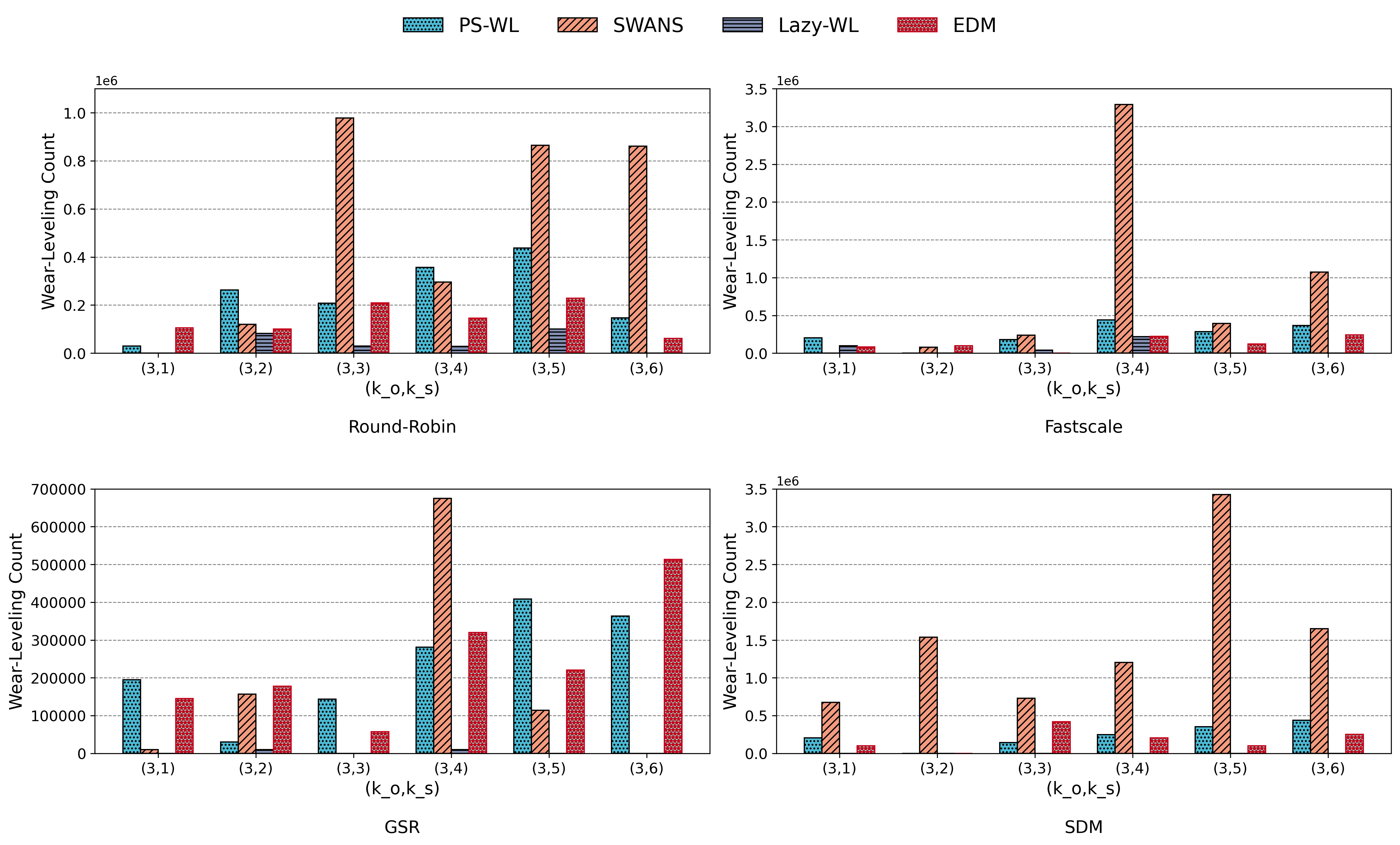}
\caption{Wear-Leveling Trigger Count result when adding scaling disks}
    \label{fig:exp1_wl_alibaba_multiple}
\end{figure}

Among these schemes, SWANS lacks a module to restrain wear-leveling operations, leading it to trigger the most wear-leveling actions in many configurations, far exceeding other schemes. The trigger counts for Lazy-WL and EDM exhibit significant fluctuations, causing their wear-leveling control to be overly aggressive in some configurations while appearing inactive in others. On average, PS-WL improved upon SWANS, Lazy-WL, and EDM by 73.9\%, 34.4\%, and 13.8\%, respectively.

The results from the evaluation metrics above lead to the conclusion that PS-WL does not merely pursue the lowest short-term data migration cost. Instead, by executing moderate and highly efficient migration operations, it achieves a degree of wear balance far superior to other schemes, thereby maximizing the long-term reliability and lifespan of the entire SSD array. This effectiveness-oriented, adaptive overhead control strategy is one of the core advantages of the PS-WL scheme over the other baseline approaches.

\subsubsection{Total I/O Count}

The Total I/O metric records the total number of I/O operations required by different wear-leveling schemes to achieve a state of wear balance among the drives in the array under various scaling scenarios. This metric measures the efficiency of wear-leveling; a lower Total I/O count indicates that a state of wear balance is reached more quickly. To validate the effectiveness of our proposed scheme, the experiment uses different initial wear states and compares the speed at which the exit condition for the array's wear-leveling mechanism is met. This serves to evaluate PS-WL's adaptive balancing speed when facing different initial conditions and to verify its contribution to enhancing overall array reliability.

As observed in the preceding comparative experiments, the schemes based on SWANS, Lazy-WL, and EDM fail to converge wear-leveling in all configurations across multiple scenarios. Therefore, comparing the Total I/O count across schemes in a context of unlimited wear-leveling has little reference value. Consequently, to demonstrate PS-WL's adaptive capability for SSD array scaling under varying initial wear conditions, the evaluation of this metric employs an ablation study comparison scheme. This ablation scheme includes all modules except the effective lifetime model, thereby allowing for an assessment of the effectiveness of the effective lifetime concept.

To simulate the scaling of a long-use SSD array in a real-world scenario, the experiment was set up with the original disks having an average initial failure probability of 0.01\%. Using a write-intensive workload, we tested the number of I/O operations required to reach an array-level state of wear balance, as well as the Array Failure Probability at that point. This latter metric evaluates the long-term benefits of the scheme from a reliability perspective and is defined as the probability of any single disk in the array failing, which would represent data loss or the need for data reconstruction.

\begin{figure}[!htb]
    \centering
    \includegraphics[width=1\textwidth]{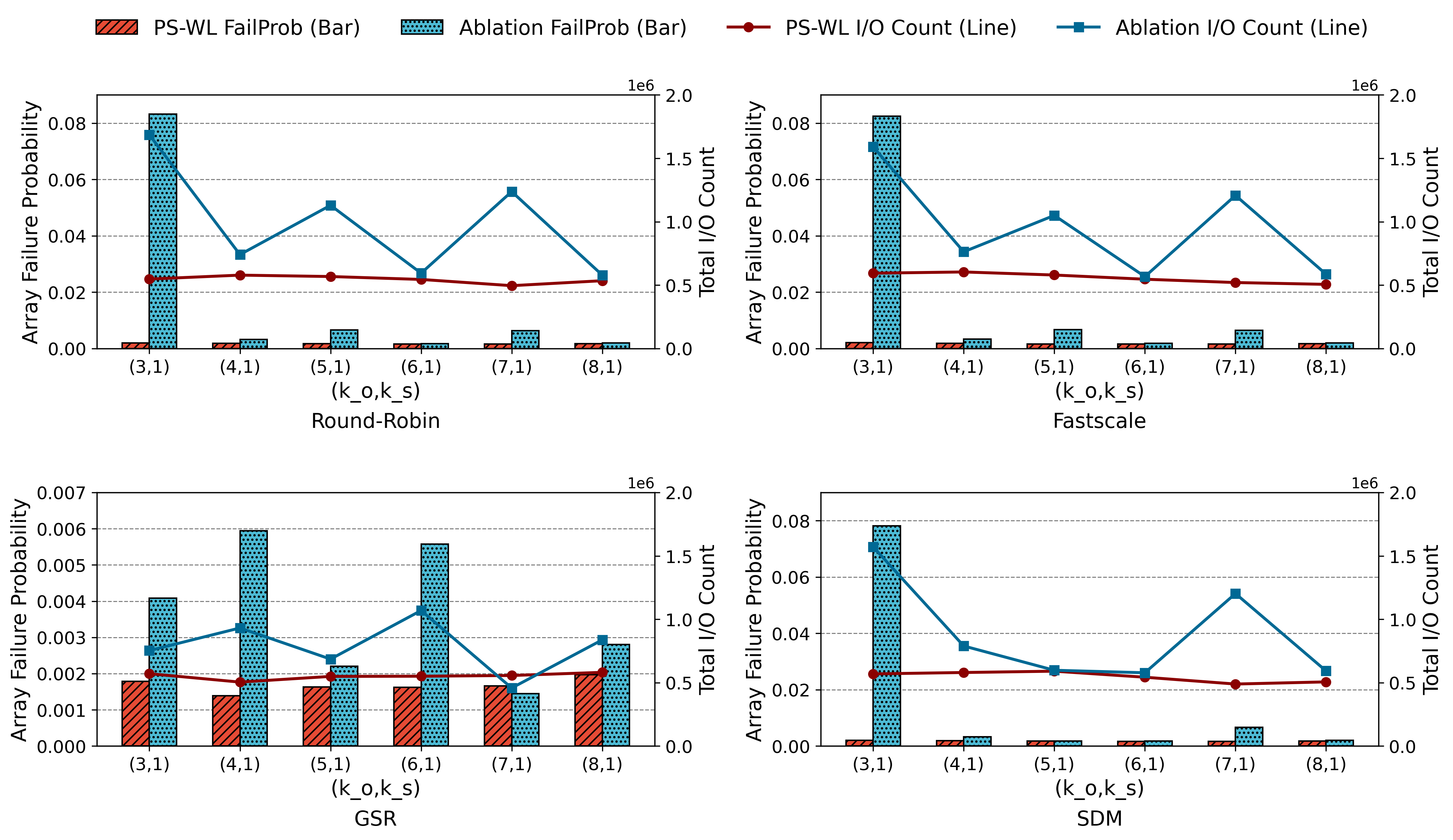}
    \caption{Observing total I/O counts of SSD array with different initial wear, under different scaling configurations in write-intensive workload}
    \label{fig:exp3}
\end{figure}

For algorithms represented by Lazy-WL and our ablation experiment, which use wear count differences as the control model input, their wear-leveling speed control remains unchanged under the same initial conditions of average wear disparity. However, as shown in Figure \ref{fig:exp3}, under high initial wear conditions, the PS-WL scheme, using the effective lifetime model, can respond as the failure probability of SSDs in the array gradually increases. It triggers more wear-leveling operations to complete the wear catch-up process more rapidly.

The results reveal a critical difference in balancing efficiency. In terms of Total I/O (the line graph), the ablation scheme (blue line) required significantly more I/O operations to achieve wear balance than the complete PS-WL scheme (red line) across all data redistribution strategies and scaling configurations. This gap was particularly pronounced in certain setups. For instance, under the Fastscale scaling scheme with a configuration of three original disks and one scaling disk, the ablation scheme executed approximately 1.63 million I/Os to meet the wear catch-up exit condition, whereas PS-WL required only about 570,000 I/Os.

The data from the bar chart further corroborates this analysis, presenting a contrast in reliability. The ablation scheme is not only inefficient but also poses a higher reliability risk. Due to the massive volume of additional writes it performs, the overall array failure probability for the ablation scheme (blue bars) is much higher than that of the PS-WL scheme (red bars) upon reaching a balanced state. By leveraging its precise perception of effective lifetime, PS-WL can achieve the wear-balancing goal with the minimum number of migrations and the least wear cost. Its wear-leveling strategy can "protect" disks with increasing failure probabilities, thereby maintaining the entire array's failure risk at an extremely low level.

In summary, this ablation study proves the effectiveness of the proposed failure-probability-based effective lifetime model. This core innovation enables PS-WL to overcome the limitations of traditional P/E count-based balancing. When handling the scaling of SSD arrays with high initial wear, it simultaneously achieves a lower balancing cost (fewer I/Os) and a higher balancing quality (lower failure probability), maintaining a low array-level failure probability over a longer period and demonstrating clear advantages in both efficiency and reliability.

\section{Conclusion}

This paper introduced PS-WL, a novel wear-leveling scheme engineered to address the critical challenges of reliability and performance in scalable SSD arrays. PS-WL advances beyond traditional wear metrics by shifting the focus from simple P/E cycle counts to a fine-grained, real-time failure probability model. This intelligence, coupled with an adaptive PID controller, provides the adaptability and robustness required to manage diverse RAID configurations and dynamic system workloads.

Comprehensive simulation results demonstrate that the PS-WL scheme exhibits significant advantages across multiple dimensions. Compared to existing schemes, it substantially optimizes the wear-leveling performance and response time of SSD arrays under various scaling configurations and test conditions, effectively extending the overall lifespan of the system. This validates its feasibility and advancement in both theoretical design and practical application.

\bibliographystyle{unsrt}  
\bibliography{references}

\end{document}